\documentclass{PoS}

\makeatletter
\newif\if@restonecol
\makeatother

\usepackage[boxed]{algorithm2e}
\usepackage{url}

\title{QCD on GPUs: cost effective supercomputing} 

\ShortTitle{QCD on GPUs: cost effective supercomputing}

\author{\speaker{M. A. Clark}$^{ab}$\\
\llap{$^a$}Harvard-Smithsonian Center for Astrophysics, 60 Garden St, Cambridge,  MA 02138, USA
\llap{$^b$}Initiative in Innovative Computing, Harvard University School of Engineering and Applied Sciences, 29 Oxford St, Cambridge,  MA 02138, USA\\
E-mail: \email{mikec@seas.harvard.edu}}

      \abstract{The exponential growth of floating point power in
        graphics processing units (GPUs), together with their low
        cost, has given rise to an attractive platform upon which to
        deploy lattice QCD calculations.  GPUs are essentially many
        (O(100)) core chips, that are programmed using a massively
        threaded environment, and so are representative of the future
        of high performance computing (HPC).  The large ratio of raw
        floating point operations per second to memory bandwidth that
        is characteristic of GPUs necessitates that unique algorithmic
        design choices are made to harness their full potential.  We
        review the progress to date in using GPUs for large scale
        calculations, and contrast GPUs against more traditional HPC
        architectures.}

\FullConference{The XXVII International Symposium on Lattice Field Theory - LAT2009\\
		 July 26-31 2009\\
		 Peking University, Beijing, China}

\begin{document}

\section{Introduction}

The lucrative gaming market has led to the exponential growth in the
floating point performance of graphics processing units (GPUs) which
has far outstripped the increase in performance of traditional CPUs.
Coupled with a similarly increasing memory bandwidth, GPUs represent a
very attractive platform upon which to deploy lattice QCD
calculations.  The combination of CPU with GPU represents an example
of a {\it heterogeneous architecture}, which is rapidly becoming the
norm for high performance computing.

This article focuses on how to utilize GPUs effectively for QCD
calculations, reviews the status quo and looks to future opportunities
for using GPUs.  The outline is as follows: in \S\ref{sec:gpu} we
review current GPUs, in particular the CUDA platform,
\S\ref{sec:dirac} describes how to effectively implement the action of
the Dirac operator upon GPUs, \S\ref{sec:inverter} focuses upon mixed
precision solvers, \S\ref{sec:multi-gpu} considers prospects for
multiple GPU parallelization, \S\ref{sec:who} reviews effective price
performance of current GPU solutions, and in \S\ref{sec:conclusions}
we present our conclusions.

\section{Graphics Processing Units}
\label{sec:gpu}

There are currently two companies highly vested in the high
performance GPU market: AMD~\cite{amd} and NVIDIA~\cite{nvidia}.  Both
companies offer highly programmable GPUs, and their current top of the
range products offer similar bandwidth to their respective memories.
Historically, programming GPUs required the use of graphics APIs,
e.g., OpenGL, which are not suitable for deploying typical scientific
applications.  In a pioneering work by Egri {\it et al}
\cite{Egri:2006zm} lattice QCD was shown to map well to GPU
architectures, obtaining exceptional price performance.  However, it
was not until the introduction of NVIDIA's CUDA (Compute Unified
Device Architecture) platform which lowered the barrier to GPU
computing that resulted in an explosion of interest in harnessing GPUs
for lattice QCD calculations~\cite{Barros:2008rd, Ishikawa:2008pf,
  Chiu:2009wh, Clark:2009wm}.  Since almost all current focus is upon the
CUDA platform, we shall here on in consider NVIDIA's CUDA architecture
only, except where noted.

\begin{table}[htb]
\begin{center}
\begin{tabular}{|l|c|c|c|c|c|c|} \hline
&  &  & GBs\(^{-1}\) & \multicolumn{2}{|c|}{Gflops} &  GiB \\ \hline 
Card & Generation & Cores & Bandwidth & 32-bit & 64-bit & Device RAM  \\ \hline
GeForce 8800 GTX & G80 & 128 & 86.4 & 518 & -  & 0.75 \\ \hline
Tesla C870            & G80 & 128 & 76.8 & 518 & -  & 1.5 \\ \hline
GeForce GTX 280   & GT200 & 240 & 142  & 933 & 78 & 1.0 \\ \hline
Tesla C1060          & GT200 & 240 & 102  & 933 & 78 & 4.0 \\ \hline
Tesla C2070          & Fermi & 512 & \(\sim200\) &  1260 & 630 & 6.0 \\ \hline
\end{tabular}
\caption{\label{table:specs}Specifications of representative NVIDIA
  graphics cards.  The G80 was the first GPU architecture to support
  CUDA, GT200 is the current generation which introduced double
  precision and Fermi is the next generation due for release early in
  2010.  The GeForce range represent the consumer gaming cards, and
  Tesla is the professional range which typically have a significantly
  increased device memory (at a significant price
  premium)~\cite{tesla-fermi}.}
\end{center}
\end{table}

GPUs are typified by massive single precision floating point
performance and a very wide, and hence fast, bus to their on card
memory (here on referred to as device memory).  They consist of
hundreds of cores (called stream processors) grouped together in many
multiprocessors.  With each passing generation the number of cores has
historically doubled, and together with a widening of the memory
interface, this has lead to GPUs quickly outpacing the performance of
traditional CPUs (Table \ref{table:specs}).  

\begin{figure}[htb]
\begin{center}
\includegraphics[width=.7\textwidth]{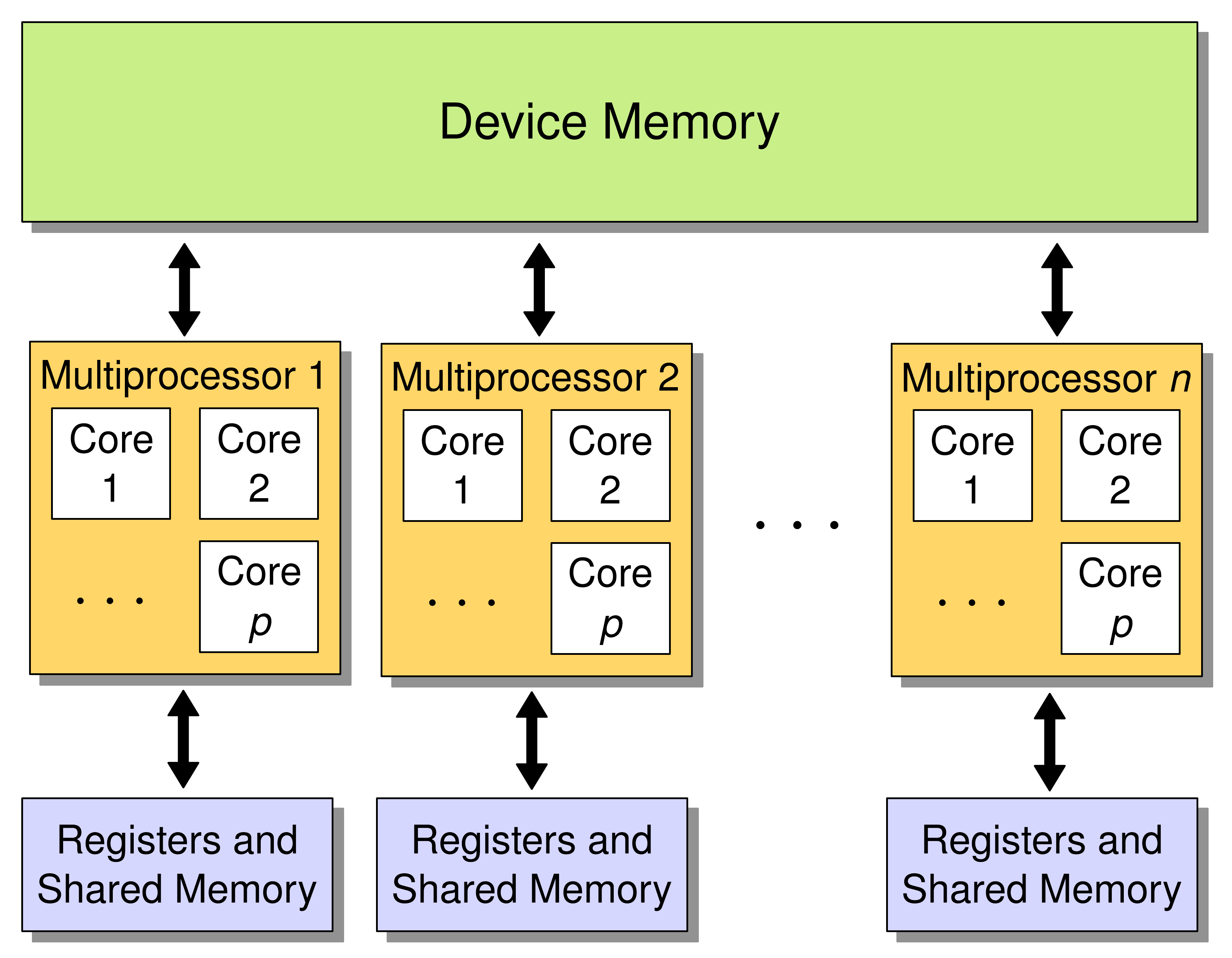}
\caption{\label{fig:arch}Architecture of a modern NVIDIA graphics
  card.  In NVIDIA's nomenclature, cores called {\it stream
    processors} (or {\it scalar processors}), and in the GT200
  generation each multiprocessor has eight such
  cores, 64 kB of registers and 16 kB of shared memory.}
\end{center}
\end{figure} 

Current generation GPUs lack a traditional cache, and so hide
latencies by running many threads concurrently:\footnote{768 threads
  per multiprocessor on current hardware.} if a group of threads
encounters a stall due to memory latency, a new group of threads will
be swapped in their place and run until either completion or indeed
until they stall.  GPUs are designed to run thousands of threads
concurrently, and although they have a very large register file, in
addition require fined grained parallelism to be fully exploited.  The
{\it occupancy} represents the fraction of the maximum number of
possible threads that is achieved by a given {\it kernel} (a function
that runs on the GPU).  In general, a greater occupancy will lead to
greater performance.  In addition, each multiprocessor has a small
user managed cache, the shared memory, this allows communication
between the cores within a multiprocessor.  The size of the device
memory varies from model to model: current gaming (GeForce) cards have
typically 1 GB on board whereas the professional Tesla cards have up
to 4 GB on board and are available in various form factors suitable
for high performance computing.  In order to realize the peak
bandwidth to device memory, reads and writes from consecutive threads
on a given multiprocessor must access adjacent regions of memory, such
transfers are known as being {\it coalesced}.  For full coalescing
each thread in a group of 16 threads must read data in either 32, 64
or 128 bit consecutive chunks.  Although GPUs have extremely large
memory bandwidth, the peak single precision floating point capacity to
device memory bandwidth ratio means that they are extremely bandwidth
limited and so efficient use of the registers and shared memory to
minimize communication to device memory (even at the expense of
introducing additional floating point work) is of paramount importance
to obtain high performance.

Double precision capability to GPUs was introduced with NVIDIA's GT200
generation, however its use invokes a large penalty relative to single
precision.  This has motivated the use of mixed precision methods,
specifically linear solvers, to circumvent this
defect~\cite{Egri:2006zm, Chiu:2009wh, Clark:2009wm}.  The next generation
architecture from NVIDIA, {\it Fermi}~\cite{fermi}, will address this
issue (as as well as introducing ECC protection) with only a factor of
two difference between the peak single and double precision floating
point arithmetic rates (Table \ref{table:specs}).

The proverbial elephant in the room is the PCI express bus, with
typical bandwidth 5 GB\(^{-1}\), through which all communication with
the host CPU must take place.  Minimizing communication through this
bus is of utmost importance in order to maximize performance.  This
typically requires that the GPU is not used as an {\it accelerator}
for a single function, rather the complete algorithm must be deployed
on the GPU.

The most popular programming interface to CUDA is the C for CUDA
interface.  This presents the application programmer with a massively
threaded C-like programming language from which the underlying
graphics hardware can be effectively exploited.  From a programming
point of view, threads are grouped together in thread blocks,
consisting of a minimum of 32 threads (though at least 64 are required
for optimum performance), and thread blocks are arranged in a grid.
Each of these thread blocks will run on a multiprocessor, and ideally
(for optimum latency hiding) multiple thread blocks will be running
concurrently on a multiprocessor.  Communication between threads in a
thread block is possible through the shared memory, however, no
communication is possible between different thread blocks without
going back to device memory (hence a global synchronization).
Branching is possible by different threads in a thread block, however,
if a group of 32 threads (called a Warp) diverge on the branch
condition, the different routes of the branch will be executed serially
potentially drastically affecting performance.

While NVIDIA has the most momentum in GPU computing with CUDA, a new
multi-platform standard called OpenCL~\cite{opencl}, is intended to
level the playing field: this brings C for CUDA like abstractions and
compilers are available for both multi-core CPUs as well as both
NVIDIA and AMD GPUs~\cite{amd-opencl}.  At the time of writing OpenCL
is still a developing standard that lacks the maturity of CUDA,
however, this will potentially be an important API in the future to
allow for multi-platform deployment with minimal code development.

\section{Dirac Operator}
\label{sec:dirac}

To date the focus has been upon implementing the most time consuming
part of lattice QCD to GPUs: the linear equation solve of the Dirac
operator.  As noted in the previous section, the entire solver must be
deployed on the GPU to prevent the onset of Amdahl's law.  Specific
issues relating to the solver will be covered in \S\ref{sec:inverter};
here we concentrate on the implementation of the Dirac operator.

As mentioned above, GPUs require fine grained parallelism for optimum
performance.  For any grid based computations, this typically equates
to assigning a single thread to each point in the grid.  This is
equally true for the application of the Dirac operator to the
spacetime lattice, where typically a single thread is assigned to
updating a single spacetime point.\footnote{As a gather operation,
  since a scatter formulation would cause a race condition between
  threads.}  Thus for a lattice of size \(N_xN_yN_zN_T\), a grid of
thread blocks with this total number of threads are created to each
update their assigned lattice site.  High register pressure will
reduce the number of concurrent threads, so judicious use of both
shared memory and registers is required to keep the occupancy
high.\footnote{A strategy that has not yet been explored is finer
  grained parallelization within either color, spin or complexity.
  Such extreme parallelization will likely be required on future
  architectures since the number of cores is expected to grow faster
  than the number registers and shared memory.}  Since each site or
link matrix consists of many entries, care must be taken with regards
to field ordering in order to obtain coalesced memory access: fields
must be reordered with internal degrees of freedom (e.g., color, spin)
running slowest since each adjacent thread can read at most 128
bits~\cite{Egri:2006zm}.  Some variations on these themes are
possible, these will be mentioned where relevant below.

\subsection{Wilson}

The application of the Wilson dslash operator requires that, for every
site in the lattice, we load the eight neighbouring sites (24 real
numbers, forwards and backwards in four dimensions) and the link
matrices connecting these sites (18 real numbers), perform the SU(3)
matrix-vector multiplication and save the resultant (24 real numbers).
For the even-odd preconditioned operator this equates to 1368 flops
(using spin projection) and 1440 bytes of memory traffic (in single
precision) per site.  Given the ratio of peak single precision flops
and the bandwidth to device memory, it is clear that this is a memory
bandwidth bound operation.  Thus to improve performance it is clear
that any reductions in memory traffic will lead to higher throughput.
Memory traffic reduction strategies for the Wilson-Dirac operator are
thoroughly explored in~\cite{Clark:2009wm}:

\begin{enumerate}
\item It is conventional in the lattice QCD community to use the
  DeGrand-Rossi basis (a chiral basis) for the Dirac matrices.  In
  this basis, the spin projectors \(P^{\pm\mu}\) for each of the 4
  dimensions have 2 non-zeros per row.  By changing to the
  non-relativistic basis, which diagonalizes \(\gamma_4\),
  \(P^{\pm4}\) has only 2 non-zeros in the entire matrix; only a half
  spinor need be loaded when fetching the neighbouring sites in the
  temporal dimension, e.g.,
\[P^{+4} = 
\left(\begin{array}{rrrr}
  1&0&\pm1&0\\
  0&1&0&\pm1\\
  \pm1&0&1&0\\
  0&\pm1&0&1\\
\end{array}\right) \Longrightarrow
P^{+4} = 
\left(\begin{array}{rrrr}
  2&0&0&0\\
  0&2&0&0\\
  0&0&0&0\\
  0&0&0&0\\
\end{array}\right),\,
P^{-4} = 
\left(\begin{array}{rrrr}
  0&0&0&0\\
  0&0&0&0\\
  0&0&2&0\\
  0&0&0&2\\
\end{array}\right).
\]

\item In a similar vein, a gauge transformation can be applied to
  rotate all the links in one dimension to the unitary matrix
  (excluding a boundary term).  Doing so means that the gauge field in
  this dimension need not be loaded (nor the connected spinor
  multiplied).

\item Given that \(U \in SU(3)\), the link matrices can be simply
  parametrized using 12~\cite{De Forcrand:1986af} or
  8~\cite{Bunk:1985rg} numbers (\(=N_c^2 - 1\) the number of
  generators of the group).  Memory traffic can therefore be reduced
  by loading the parametrization of \(U\) and reconstructing the full
  matrix once in registers.  Using such parametrizations increases
  the total number of operations to apply the Wilson operator by 384
  flops (25\%) and 856 flops (63\%), however, as is a common theme in
  GPU computing, reducing memory traffic at the expense of floating
  point operations increases overall throughput.  Using a
  parametrization for the gauge field has the added benefit of
  reducing the memory footprint, which can be vital on the GPU where
  memory is at a premium.
\end{enumerate}

\begin{figure}[htb]
\begin{center}
\includegraphics[width=.7\textwidth]{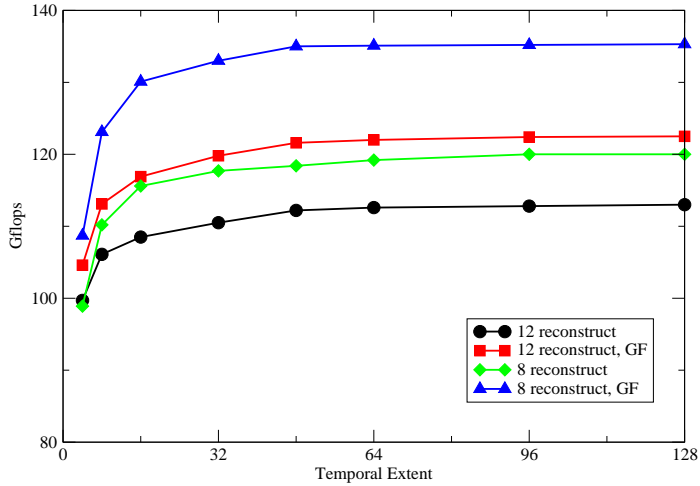}
\caption{\label{fig:wilson}Performance of the single precision
  even-odd preconditioned Wilson-Dirac matrix-vector product on a GTX
  280 as a function of temporal length (spatial volume \(24^3\); GF
  denotes that temporal gauge fixing has been
  applied)~\cite{Clark:2009wm}.}
\end{center}
\end{figure}
 
Figure \ref{fig:wilson} is a plot demonstrating the single precision
performance using various combinations of the strategies highlighted
above.\footnote{The reported Gflops are effective Gflops, i.e., not
  including the cost of the \(SU(3)\) matrix reconstruction, nor the
  savings from imposing gauge fixing.}  The conclusion is clear:
reducing memory traffic leads to higher performance.  Note that at
smaller volumes the overall performance is reduced because there is
less parallelism, and hence less threads to hide latency.

In double precision, because of the much lower floating point peak
performance (on the GT200 generation), the flop / bandwidth ratio is
now more equally balanced.  As such there is actually a net reduction
in performance using the more flop intensive 8 parameter gauge field
method, and the simpler 12 parameter strategy is preferred.  The ratio
between single and double precision actual performance is typically
only a factor of 3-4 (see Table \ref{table:invert-perf}).

Another memory traffic reduction strategy is to decrease the precision
in which the spinor and gauge fields are stored.  In particular, the
use of half precision for reading and writing to device memory (while
still performing the computation in single precision) halves the
memory traffic, corresponding to an almost doubling in performance
(see Table \ref{table:invert-perf}).  When combined with a mixed
precision solver, this can lead to a net reduction in the time to a
double precision solution (see \S\ref{sec:inverter}).

\subsection{Wilson-Clover Fermions}

With a high performance Wilson dslash implemented, the extension to
Wilson-Clover is straightforward.  It is worth pointing out, however,
that in the non-relativistic Dirac basis the clover term is a fully
dense matrix, hence the approach taken in the QUDA library~\cite{QUDA}
is to convert the spinor field into the relativistic basis on the fly
(and back again after the application of the clover term) to avoid
incurring the extra bandwidth and storage.  Performance is typically
slightly greater (\(\sim 10\%\)) than the Wilson operator as the
addition of the clover term increases the overall compute intensity.

\subsection{Domain Wall Fermions}

In the work by Chiu {\it et al} \cite{Chiu:2009wh} they describe how
to implement a \(5d\) formulation of chiral fermions for CUDA.  Here
they use \(4d\) even-odd preconditioning which results in the same
gauge field along all sites of the \(5^{th}\) dimension (of length
\(N_s\)).  Thus, by sharing the gauge field load across all threads
acting on a given \(4d\) point and storing these elements in shared
memory the memory traffic for the gauge field can be reduced by
\(O(1/N_s)\).  In addition they create a grid of thread blocks
consisting in total of only \(N_x N_y N_z N_s\) threads, and each one
these threads streams through the temporal dimension reusing the
forward hop spinor from the previous site as the backward hop spinor
for the current site.  The resulting CG inverter performance is 136
Gflops sustained in single precision on a GTX 280, and they use a
mixed precision approach (see \S\ref{sec:inverter}) to achieve double
precision accuracy which reduces the overall performance to 120 Gflops
sustained.

\subsection{Overlap Fermions}

\begin{figure}[htb]
\begin{center}
\includegraphics[width=.35\textwidth]{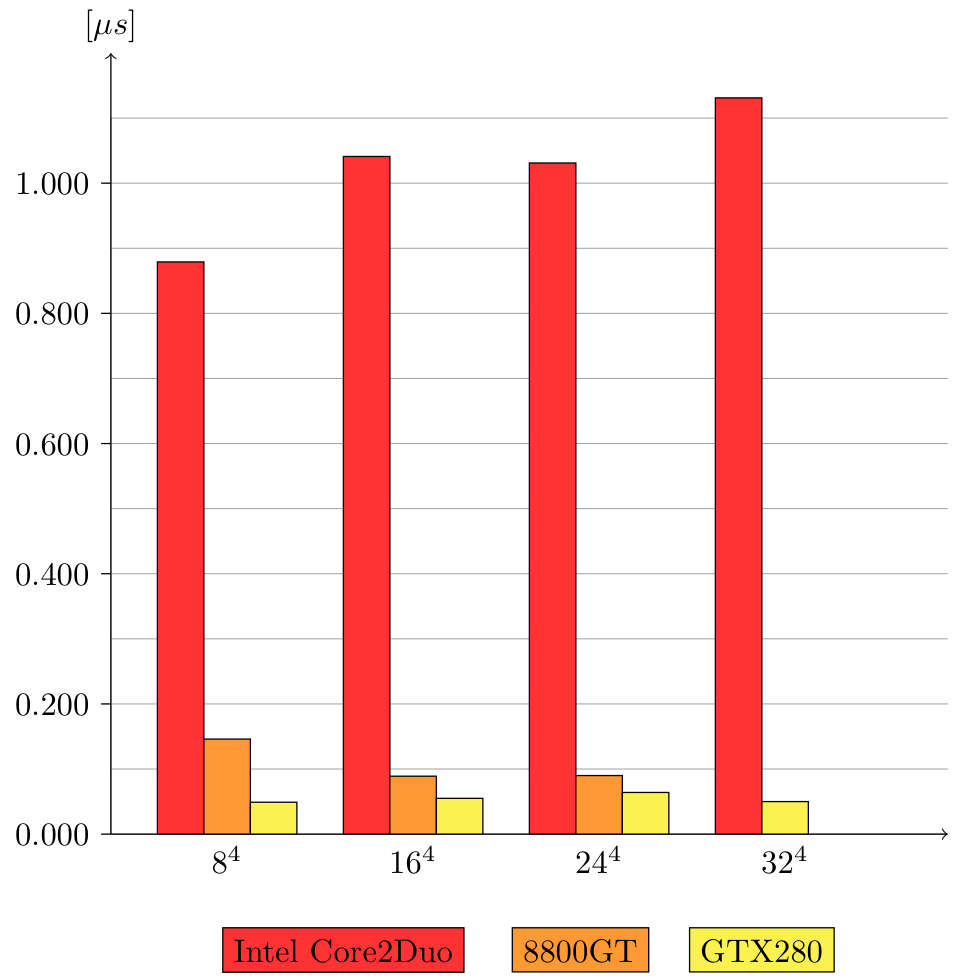}
\includegraphics[width=.35\textwidth]{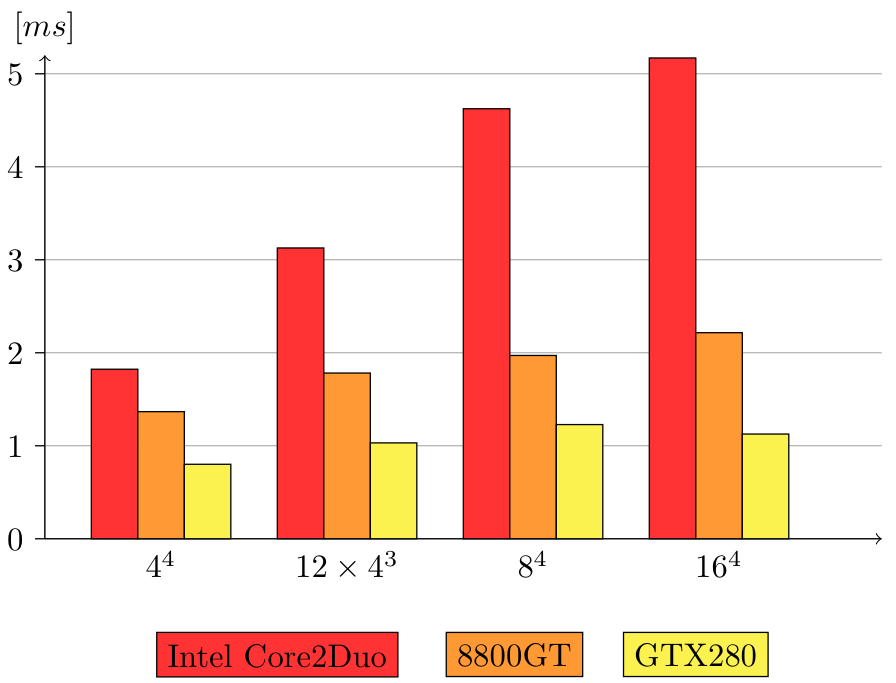}
\caption{\label{fig:overlap} Left panel: runtime of the application of
  the Neuberger operator without low-modes.  Right panel: runtime of
  low-mode solver.  The runtime is normalized to the lattice volume
  and the degree of of the approximating polynomial (left) and the
  number of eigenvalues found (right).}
\end{center}
\end{figure}

With a Wilson kernel implemented, the application of this to the
Neuberger operator is a natural extension.  On traditional
architectures the sign function \(\epsilon(H)\) is usually best
approximated using an optimal rational approximation (Zolotarev)
evaluated using a multi-shift solver.  However, such solvers map
poorly onto GPUs because of limited memory and the bandwidth intensive
nature of the additional linear algebra would significantly affect
performance.  The strategy taken by Wittig and Walk~\cite{Wittig} is
to evaluate \(\epsilon(H)\) on the GPU, and run the outer inverter on
the host.  Here \(\epsilon(H)\) is approximated using a Chebyshev
polynomial using the Clenshaw recurrence relation: such an approach is
well suited to the GPU because of small memory and linear algebra
overhead.  In single precision, they achieve 65 Gflops on a GTX 280,
which represents a 20 fold speed up over their CPU implementation
(left panel of Figure \ref{fig:overlap}).  To find the low modes
required for projection, they are using the GPU simply as an
accelerator for the eigenvector solver which runs on the host CPU, and
as such the speedup over the CPU is limited (right panel of Figure
\ref{fig:overlap}).

\subsection{Staggered Fermions}

Obtaining high performance for staggered fermions on GPUs is more
difficult than for Wilson fermions because of the decreased
compute intensity, and since many improved staggered fermion
variants use link matrices that are smeared such that \(U \not\in SU(3)\), meaning that no bandwidth reducing
parametrizations are possible.  Staggered fermions do have the
advantage of involving less degrees of freedom, and so a larger
spacetime lattice will fit on a single GPU.

A number of groups are using staggered fermions with GPUs: Cossu {\it
  et al} \cite{Cossu} are exploring the \(N_f=2\) staggered fermion
phase transition for a variety of small masses exploiting trivial
parallelism to explore the parameter space.  On the Tesla C1060 they
achieve 60 Gflops in single precision for the complete CG inverter,
this represents a 50 fold speed up over the CPU.  However, when they
consider HMC, only a 20 fold speedup is obtained: this is the onset of
Amdahl's caused by the gauge force and momentum updates being done on
the CPU and hence slowing down the entire calculation.  To rectify
this the entire HMC trajectory must be done on the GPU.  In addition
Shi~\cite{guochun} has ported variations of a staggered CG solver to
the QUDA package~\cite{QUDA} and observes 30, 86 and 135 Gflops for
double, single and half precisions respectively on a GTX 280.

\section{Mixed Precision Solvers}
\label{sec:inverter}

With the action of the Dirac operator upon a spinor field implemented,
the implementation of a Krylov solver is straightforward.  The
additional linear algebra required can be found in the cuBLAS library,
(included with the CUDA SDK) and provides a mostly complete
implementation of BLAS~\cite{blas}.  However, since these operations
are extremely bandwidth bound it is better to fuse such linear algebra
operations into single kernels where possible to reduce memory
traffic.  Table \ref{table:invert-perf} lists some typical performance
numbers for CG and BiCGstab solvers at different precisions on the GTX
280, where it can be seen that actual solver performance is typically
15\% slower than the equivalent ``bare'' kernel operation because of
the additional linear algebra.

\begin{table}[htb]
\begin{center}
\begin{tabular}{|l|c|c|c|c|}\hline 
Kernel type & Precision & Kernel (Gflops)& CG (Gflops) & BiCGstab
(Gflops) \\ \hline 
Wilson  & Half & 207.5 & 179.8 & 171.1 \\ \hline 
Wilson & Single & 134.1 & 116.1 & 109.9 \\ \hline 
Wilson & Double & 40.3 & 38.3 & 37.9 \\ \hline
\end{tabular}
\caption{\label{table:invert-perf}Performance comparison of the Wilson
  even-odd matrix-vector kernels with the associated CG and BiCGstab
  solvers on the GeForce GTX 280 (volume = \(24^3\times48\), double
  precision global sums)~\cite{Clark:2009wm}.}
\end{center}
\end{table}

Given the disparity in performance at different precisions, and that
typically greater than single precision accuracy is required when
solving the Dirac equation, mixed precision solvers have become a {\it
  de facto} tool when using GPUs.\footnote{Mixed precision solvers are
  increasingly used on more traditional architectures such as
  BlueGene~\cite{bagel}.}  The standard approach is defect-correction
(also known as iterative refinement)~\cite{Wilkinson:1966} and allows
the residual to be reduced in an inner solve using low precision,
while the residual calculation and solution accumulation are done in
high precision (see Algorithm \ref{alg:defect}).  Such an approach is
guaranteed to converge to the desired precision \(\epsilon\) provided
that the inverse of the spectral radius is bounded by the unit of least
precision of the arithmetic used for the inner solve, i.e.,
\(1/\rho(A) > \mbox{ulp}^{in}\), where \(A\) is the system matrix in
question.

\begin{algorithm}[H]
\SetLine
\(r_0\) = \(b - Ax_0\)\;
\(k\) = 0\;
\While{\(||r_{k}|| > \epsilon\)}{
\mbox{Solve} \(A p_{k+1}\) = \(r_{k}\) \mbox{to precision} \(\epsilon^{in}\)\; 
\(x_{k+1}\) = \(x_k + p_{k+1}\)\;
\(r_{k+1}\) = \(b - Ax_{k+1}\) \;
\(k = k + 1\)\;
}
\label{alg:defect}
\caption{Defect-correction solver for \(Ax=b\) (initial guess
  \(x_0\), outer solver tolerance \(\epsilon\) and inner solver
  tolerance \(\epsilon^{in}\)).}
\end{algorithm}

The disadvantage of using defect-correction is that each time the
residual is recalculated in high precision, the low precision Krylov
solver is restarted, thus losing the orthogonal sub-space that has
been built up; thereby increasing the total number of iterations to
reach convergence.  The use of {\it reliable updates}
\cite{Sleijpen:1996} in the context of a mixed precision solver was
introduced in \cite{Clark:2009wm}, as a means to circumvent the
restarting penalty.  Here, the residual is recalculated and the
solution accumulated in high precision more frequently, however, this
is done {\it in situ}, without an explicit restart (see Algorithm
\ref{alg:reliable}).  Although this makes little difference when
compared to defect-correction in a single-double mixed precision
solver, for half-double this was found to significantly decrease the
number of iterations until solution~\cite{Clark:2009wm}.

\begin{figure}[htb]
\begin{center}
\includegraphics[width=.7\textwidth]{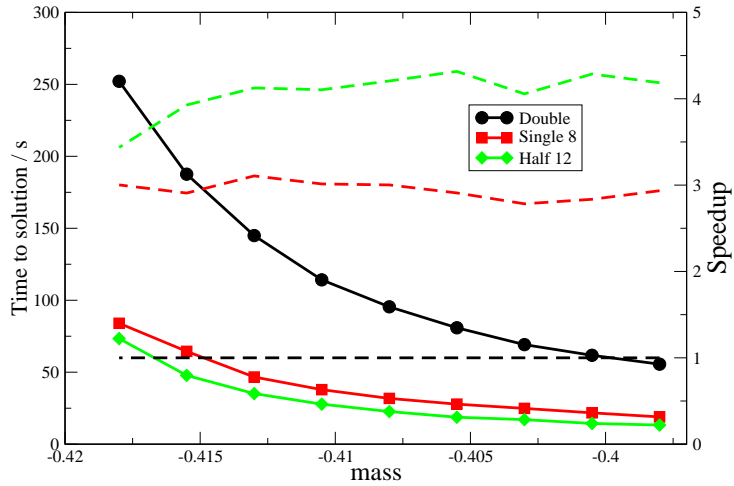}
\caption{\label{fig:time} Time to solution (solid lines) for double
  precision and reliable update solvers and speedup versus double
  precision (dashed lines) (BiCGstab, \(\delta=0.1\),
  \(\epsilon=10^{-12}\), volume = \(24^3\times
  64\))~\cite{Clark:2009wm}.}
\end{center}
\end{figure}

Of course the only metric that is of real concern is the time to an
accurate solution.  Figure \ref{fig:time} is a plot comparing the time
to solution of using a pure double precision solver with single and
half precision solvers with double precision reliable updates.  The
desired final residual tolerance \(\epsilon=10^{-12}\) is far the
beyond the resolution of single precision.  There is a significant
reduction in the time to solution when using a mixed precision
approach.  Also shown is the speedup versus the pure double solver:
here the speedup using single precision is around a factor 3, and for
half precision it is a factor of 4.  The decrease in speedup as the
chiral limit is approached is caused by the inability half precision
to correctly resolve the full eigenvalue spectra, but in the region of
physical interest, the half-double approach is always the fastest.

\begin{algorithm}[H]
 \SetKwData{True}{true}
\SetLine
\(r_0\) = \(b - Ax_0\)\;
\(\hat{r}_{0}\) = \(r\)\;
\(\hat{x}_{0}\) = 0\;
\(k\) = 0\;
\While{\(||\hat{r}_{k}|| > \epsilon\)}{

Low precision solver iteration: \(\hat{r}_{k}\rightarrow \hat{r}_{k+1}\), \(\hat{x}_{k}\rightarrow \hat{x}_{k+1}\)\;

\If{\(||\hat{r}_{k+1}||<\delta M(\hat{r})\)}{
\(x_{l+1}\) = \(x_{l}+\hat{x}_{k+1}\)\;
\(r_{l+1}\) = \(b - Ax_{l+1}\) \;
\(\hat{x}_{k+1}\) = 0\; 
\(\hat{r}_{k+1}\) = \(r\)\;
\(l\) = \(l+1\)\;
}
\(k\) = \(k + 1\) \;
}
\label{alg:reliable}
\caption{Mixed precision reliable update solver for \(Ax=b\) (initial
  guess \(x_0\), outer solver tolerance \(\epsilon\), \(M(r)\) is the
  maximum of the norm of the residuals since the last residual update,
  ( \(\hat{}\) ) denotes low precision)~\cite{Clark:2009wm}.}
\end{algorithm}

\section{Multi-GPU}
\label{sec:multi-gpu}

Although reasonably large lattice volumes can be accommodated for on
current GPUs, the desire to go to larger volumes and perform the gauge
generation on GPUs necessitates that the calculation is spread over
multiple GPUs.  This is the next step to be taken for QCD on GPUs, and
at the time of writing none of the QCD research groups have suitable
multi-GPU production code.

There are two alternatives for multi-GPU operation:
\begin{itemize}
\item Place many GPUs in a node, and parallelize only within the node.
\item Connect many nodes together, each with one or more GPUs, by a fast inter-connect.
\end{itemize}

Currently, it is not possible to copy data directly from GPU to GPU,
and one is forced to go through the CPU host memory.  However, it is
possible to perform asynchronous transfers, i.e., overlap the
communication between the GPU and CPU while the GPU is executing a
kernel.\footnote{In addition, the Fermi architecture introduces
  bi-directional transfers, so that the GPU can be sending and
  receiving while executing a compute kernel.}  While going through
the CPU memory does not affect the aggregate bandwidth it does
increase latency, increasing the minimum local volume feasible on each
GPU.  Likewise, when sending and receiving messages through
Infiniband, the data must be read through a CPU buffer.  The use of
threads over processes is preferred for multiple GPUs within a node to
avoid the extra message sending overhead.

\begin{figure}[htb]
\begin{center}
\includegraphics[width=.7\textwidth]{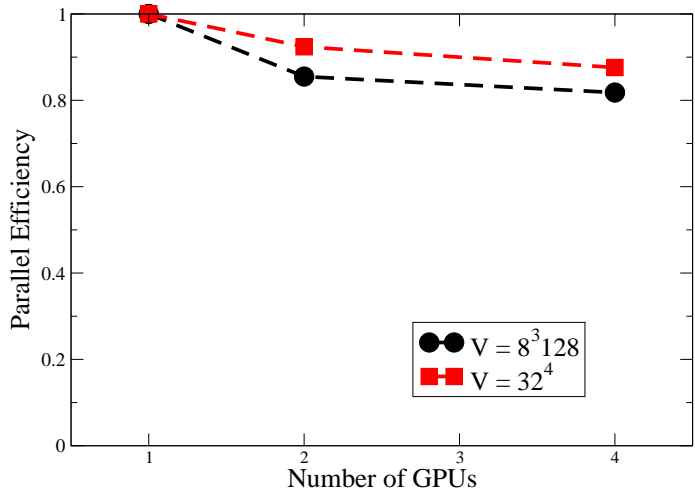}
\caption{\label{fig:multi-gpu} Weak scaling parallel efficiency of the
 Wilson Clover operator as function of number of GPUs (Tesla S1070,
 one process per GPU, temporal parallelization only)~\cite{jie}.}
\end{center}
\end{figure}

Initial attempts at limited multi-GPU operation within a node have
been extremely promising.  Even without the overlap of communication
and computation, and the overhead of using MPI instead of threads to
control each GPU, up to 90\% parallel efficiency with 4 GPUs has been
obtained (Figure \ref{fig:multi-gpu}).  Clearly however, moving beyond
the confines of a single node will be challenging, and alternative
algorithms (e.g., Schwartz approaches~\cite{Luscher:2003qa}) may be
required for parallelizing over many GPUs.

\section{Price performance}
\label{sec:who}
The current generation of GPUs attain around 100 Gflops of sustained
performance in single precision for Wilson like discretizations.  This
compares to a single rack of BG/P which sustains around 3
Tflops~\cite{bagel}.  When we compare the upfront cost (Tesla C1060
\$1200 and a host computer, BG/P a lot) and running costs (typical
GT200 power consumption is 225W and probably the same again for the
host, whereas a single BG/P rack requires 30kW~\cite{oakridge}) this
makes GPUs a very attractive proposition for QCD calculations.

When designing GPU clusters for lattice QCD calculations, the nature
of the target calculations will have an effect on the required
hardware, and hence the price performance.  For trivially
parallelizable applications (e.g., those involving many solutions to
the Dirac equation, with a constant gauge field), there is no
requirement for a fast inter-connect between the nodes.  Even for
those calculations that require the use of multiple GPUs because of
memory contraints, the most cost efficient approach is to use multiple
GPUs within a node.  As soon as one considers a multi-node cluster the
cost will increase significantly because of the added cost of a fast
interconnect, and the lower Gflops per node.

There are an increasing number of large scale GPU clusters being used
for lattice QCD calculations, e.g., Wuppertal, Thomas Jefferson
Laboratory (Jlab) and TWQCD.  These clusters have been built with the
primary goal of multi-GPU within each node only\footnote{Both the Jlab
  and TWQCD GPU clusters contain nodes totalling 32 GPUs connected by QDR
  Infiniband.} (and currently using only a single GPU per Dirac
inversion).  Both the Jlab and TWQCD GPU clusters are built using the
GT200 generation GPUs, consisting of 128 and 202 GPUs respectively (at
the time of writing).  For domain wall fermions and Wilson clover
fermions, on which these respective groups focus, both report a price
performance around \(\$0.02\) per Mflop~\cite{Chiu:2009wh, chip}.

We end this section by noting that in the era of heterogeneous
architectures and mixed precision algorithms, comparing price
performance is no longer as simple as \$ per Mflop.  Improved solver
algorithms such as inexact deflation~\cite{Luscher:2007se} and
adaptive multigrid~\cite{Brannick:2007ue, Clark:2008nh}, and their
efficient mapping onto evolving target architectures further couple
the algorithmic performance to the underlying hardware.  The only fair
comparisons will be those that take account for these variables:
seconds per HMC trajectory and such like.

\section{Conclusions}
\label{sec:conclusions}

With the heterogeneous paradigm now in full swing, the lattice QCD
community is well placed to embrace the increased computation
available through GPUs.  To take full advantage of the performance
offered by GPUs, algorithms must be redesigned and must be ported
completely else Amdahl's law will severly restrict any gains.  GPUs
additionally promise to vastly reduce the cost of entry into doing
lattice QCD.

The next generation GPU architecture Fermi from NVIDIA addresses
nearly all concerns that lattice field theorists may have had when
deploying their calculations on GPU platforms: fast double precision,
ECC protected memory, C++ support (the only remaining one being peer
to peer communication).  It should be noted that even with fast double
precision, mixed precision methods will still be important because
there will always be a factor of two difference in memory traffic
between single and double precisions.

The remaining hurdle before GPUs will hit prime time is that of
multi-GPU deployment.  While limited parallelization is achievable
with current technology, for large scale deployment (100s GPUs)
improved inter-GPU communication or algorithmic developments will be
required.

\section{Acknowledgements}

The author would like to thank the following people for supplying
data, figures and thoughts used in this work: Andrei Alexandru, Ronald
Babich, Kipton Barros, Richard Brower, Jie Chen, Ting-Wai Chiu, Guido
Cossu, Alistair Hart, Sandor Katz, Claudio Rebbi, Guochun Shi, Chip
Watson and Hartmut Wittig.  This work was supported in part by NSF
grants PHY-0427646 and PHY-0835713.

\end{document}